\documentstyle[twocolumn,epsf]{article}
\newcommand{\beq}{\begin{equation}}
\newcommand{\eeq}{\end{equation}}
\newcommand{\bea}{\begin{eqnarray}}
\newcommand{\eea}{\end{eqnarray}}

\def\nuc#1#2{\relax\ifmmode{}^{#1}{\protect\text{#2}}\else${}^{#1}$#2\fi}

\title{\large
$^7$Li(p,n) NUCLEAR DATA LIBRARY 
FOR INCIDENT PROTON ENERGIES TO 150 MEV\\
}
\vspace{1cm}
\author{S.~G.~Mashnik, M.~B.~Chadwick, H. G. Hughes, R. C. Little,
R.~E.~MacFarlane, \\
L.~S.~Waters, and P.~G.~Young\\
{\it Los Alamos National Laboratory, Los Alamos, NM 87545, USA}} 
\begin{document}
\maketitle
\noindent ABSTRACT
\vspace{0.1cm}

Researchers at Los Alamos National Laboratory are considering the
possibility of using the Low Energy Demonstration Accelerator,
constructed at Los Alamos Neutron Science Center
for the Accelerator Production of Tritium Project,
as a neutron source. Evaluated nuclear data are needed for the
p+$^7$Li reaction, to predict neutron production from thin and thick
lithium targets. In this paper we describe evaluation methods
that make use of experimental data, and nuclear model calculations, to
develop an ENDF-formatted data library for incident protons with
energies up to 150 MeV. The important $^7$Li(p,n$_0$) and
$^7$Li(p,n$_1$) reactions are evaluated from the experimental data,
with their angular distributions represented using Lengendre
polynomial expansions. The decay of the remaining reaction flux is
estimated from GNASH nuclear model calculations. This leads to the
emission of lower-energy neutrons and other charged particles and
gamma-rays from preequilibrium and compound nucleus decay processes.

The evaluated ENDF-data are described in detail, and illustrated 
in numerous figures. 
We also illustrate the use of these
data 
in a representative application
by a radiation transport simulation with the 
code MCNPX.

\vspace{0.4cm}
\noindent I. INTRODUCTION
\vspace{0.1cm}

As a part of the Accelerator Production of Tritium (APT) Project
[1], the Low Energy Demonstration Accelerator (LEDA)
has been constructed at the Los Alamos Neutron Science Center (LANSCE) [2].
LEDA is a high-current low-energy proton accelerator that can be
used to provide a source of neutrons, following proton bombardment 
on suitable targets. For instance, high-Z targets can be used to produce 
spallation neutrons. However, there is a recent interest in the 
use of a $^7$Li target, which, when bombarded with protons, 
can produce a relatively high yield of 
quasimonoenergetic
neutrons in
the forward direction via the  $^7$Li(p,n) reaction. In particular, 
this reaction may be useful to provide neutrons with energies 
near 14 MeV, for  materials testing for the fusion program.

In addition to the above application, quasimonoenergetic and broad
neutron sources are needed for research in other areas in
nuclear science and technology, {\it e.g.} Accelerator Transmutation
of Waste (ATW), radiation damage studies, medical isotope production,
and physics cross section experiments.
 
In order to assess the feasibility of using a lithium target to
produce neutrons, accurate evaluated cross section data are needed.
These data can be used to predict the neutron yield, as well as the
neutron energy and angular dependencies, from both thin and thick
targets.  (Thin targets allow the possibility of producing
quasimonoenergetic neutron sources\footnote{Though our results shown
in this paper indicate that once the incident proton energy exceeds a
few-MeV, even thin targets result in a substantial number of
lower-energy neutrons.}, whereas thicker targets produce broad-spectrum
neutron sources.)

Table-based charged-particle transport is a new feature of the MCNPX
transport code system and has been implemented for protons
in the developmental vesion [3]. Proton evaluations for
energies up to 150 MeV have been completed about a year ago for
42 isotopes [4]. The present study extends our library for $^7$Li,
not covered by [4]. A detailed report on our work may be found in [5].

\vspace{0.4cm}
\noindent II. AN OVERVIEW OF AVAILABLE $^7$Li(p,xn) DATA
\vspace{0.1cm}
 
An overview of reactions
used for the production of fast quasimonoenergetic neutrons,
including the $^7$Li(p,xn) reaction,
may be found in our report [5] and in the comprehensive review by Drosg [6].
Natural lithium consists of isotopes $^6$Li and $^7$Li with abundances 7.42
and 92.58\%, respectively. 
The $^7$Li(p,n)$^7$Be
reaction
was reviewed in 1960 by Gibbons and
Newson [7].
Useful information on kinematics and technical aspects related 
to the use of $^{nat}$Li may be found in [8].

Experimentally, the $^7$Li(p,n)  reaction was measured by several groups: 

   A) From threshold to about 8 MeV, by Meadows and Smith, using the 
Argonne Fast-Neutron Generator [9, 10];

   B) From 10 to about 20 MeV, by Anderson {\it et al.}, at Lawrence Radiation
Laboratory 
[11], and from 4.3 to 26 MeV, 
by Poppe {\it et al.} using EN tandem 
Van de Graaff accelerator 
(and AVF cyclotron, for $E_p > 15$ MeV) 
at Livermore 
[12].
Poppe {\it et al.} have been summarized in Ref. [12]
practically all measurements of this reaction before 1976;

   C) At 15, 20, and 30 MeV, by McNaughton {\it et al.}, at the Croker Nuclear
Laboratory 
[13];
   
   D) By a number of other groups (see, e.g., [14-18]),
   at proton energies
above 20 MeV.

Between 1.9   and 2.4  MeV bombarding energy, the neutrons are
monoenergetic and the reaction cross section is large.
Therefore the
$^7$Li(p,n)$^7$Be reaction has long been used as a source of neutrons
(n$_0$) at these energies [7].

   Above 2.4 MeV the first excited state of $^7$Be at 0.43 MeV may be excited, 
producing a second group of neutrons (n$_1$). However, below 5 MeV the 
zero-degree yield of these low energy neutrons is less than about 10\% 
of the 
ground-state yield, so that the usefulness of the reaction as a monoenergetic 
neutron source is only slightly impaired.

   Above 3.68 MeV, the threshold for the three-body breakup reaction
$^7$Li(p,n$^3$He)$^4$He, neutrons from this mode contribute also to the total 
neutron yield. The zero-degree neutron spectra from this mode are very
broad, of an evaporative-type form, and were measured up to $E_p = 7.7$ MeV
by Meadows and Smith [10].

Above 7.06 MeV, the threshold for the reaction
$^7$Li(p,n)$^7$Be with excitation of the second excited state of 
$^7$Be, neutrons
from this mode (n$_2$) also begin to contribute to the total neutron 
yield, but this contribution is not significant [10].
Therefore, the 
usefulness of the $^7$Li(p,n)$^7$Be reaction as a source of 
neutrons at energies 
above 5-7 MeV would depend on a particular application. Only a very good
energy resolution would allow one to separate the substantial number of
n$_1$ neutrons to obtain a monoenergetic source. However, if the application
can tolerate including both n$_0$ and n$_1$ 
neutron groups, then the reaction 
has the favorable feature of use of a cheap solid target with a forward 
angle laboratory cross section approaching 7.4 mb/sr at $E_p = 15$ MeV and
14.5 mb/sr at $E_p = 20$ MeV.
It should be noted that
although the neutron yield is large, the presence of low energy neutrons
from different reaction modes may limit the usefulness of the 
$^7$Li(p,n)$^7$Be
reaction as a monoenergetic neutron source at proton energies 
above 10 MeV [12].

The n$_0$ and n$_1$ neutron yields are strongly forward
peaked. Their measured cross section increases 
monotonically with increasing $E_p$
(see Table II in [5]). 
This is not true for the total neutron 
cross sections integrated over all emission angles, 
shown for $E_p < 26$ MeV in Fig. 1
(see [5] for a compilation of all
available data).

One can see that in contrast with the zero degree differential cross section, 
the integrated cross section decreases monotonically above 
$\sim 5$ MeV,
reflecting 
the onset of 
the strong forward peaking. Above $E_p = 7$ MeV, the total number of n$_1$ 
neutrons which leave $^7$Be in the first-excited states is always greater 
than 25\% of the number of n$_0$ neutrons which leave $^7$Be 
in the ground state, 
reaching a maximum of about 55\% at $E_p \sim 9$ MeV.

A compilation of 
presently available 
measured laboratory differential cross sections
at zero deg for the $^7$Li(p,n)$^7$Be
n$_0$ and n$_1$ (0.0 MeV and 0.43 MeV levels of $^7$Be) reactions and the    
ratio of n$_1$ to n$_0$ neutrons 
and of measured laboratory
integrated cross sections 
as functions of 
proton kinetic bombarding energy
may be found in our report [5].

\begin{figure}[h!]
\vspace*{-0.5cm}
\centerline{
\hspace*{+0.5cm}
\psfig{figure=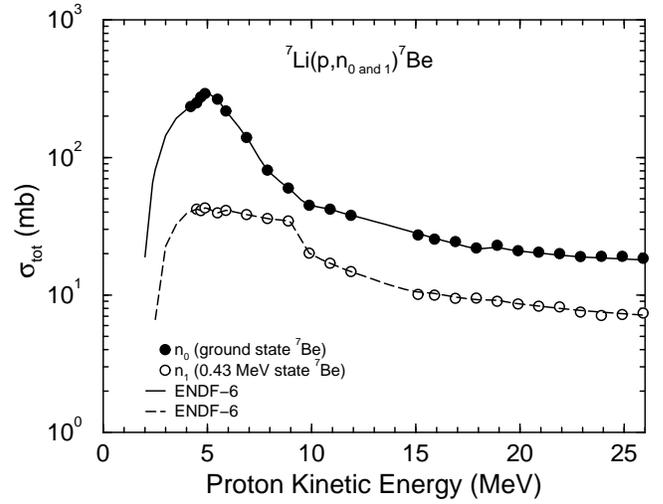,width=80mm,angle=-90}}

\caption
{
Measured total cross sections for the 
$^7$Li(p,n$_0)^7$Be
and
$^7$Li(p,n$_1)^7$Be reactions between 4 and 26 MeV [12]
together with our evaluations in the ENDF-6 [19] format.  
}  
\end{figure}

\vspace{0.4cm}
\noindent III. EVALUATION METHODS
\vspace{0.1cm}

The production of an evaluated data library for proton reactions on lithium 
poses certain difficulties. In particular, nuclear model calculations
based on statistical preequilibrium and equilibrium decay theories  
become unreliable for light target nuclei, where the nuclear levels 
are widely spaced. For this reason, one would like an evaluated data 
library that is based, as much as possible, on measured data.

We have therefore adopted the following approach. There are numerous
measurements of the important $^7$Li(p,n$_0$) and $^7$Li(p,n$_1$) direct
reactions populating the ground and first excited states of $^7$Be,
and we have used these experimental data for the cross sections and
angular distributions in the evaluation. This is important because 
many aspects of the production of neutron sources via $^7$Li(p,n) 
reactions involve use of the high-energy quasimonoenergetic n$_0$ and 
n$_1$ neutrons. 
To model the remaining 
cross section available for nuclear reactions, {\it i.e.} the overall 
reaction cross section minus these direct reaction cross sections, we 
use GNASH nuclear model calculations [20,21]. 

Use of the GNASH calculations is expected to lead to some weaknesses
in the evaluated data, because of the aforementioned difficulties in
using a statistical model code. However, we note that the calculations
do, at least, include certain important constraints such as
energy, flux, angular momentum, and parity, conservation laws. Since the
GNASH calculations make use of experimental nuclear levels
information, they can also be expected to lead to emission spectra
that correctly include peaks and gaps in the calculated
energy-dependent spectra at the correct locations, {\it i.e.} peaks at
energies where final states can be excited, and gaps where there
are no final nuclear states available.
\\

\begin{center}
A. Evaluation of $n_0$ and $n_1$ cross sections and angular 
distributions\\ 
\end{center}

The $^7$Li(p,n$_0$) and $^7$Li(p,n$_1$) evaluations were based upon
measured data. The center-of-mass measured angular distributions of
both the n$_0$ and n$_1$ neutrons were fitted using Legendre
polynomials, 
$${d \sigma _{c.m.} \over d \Omega} = \sum^{N}_{n=0}
A_nP_n(cos \theta) .$$ 

For proton incident energies up to 12 MeV, such Legendre 
fits have been published  by Poppe {\em et al.}, and therefore we 
adopted their
results (Fig. 5 of Ref. [12]).

At higher proton incident energies, 
we performed Legendre polynomial fits to measured angular distributions. 
For proton incident energies
of 15.1, 15.9, 16.9, 17.9, 18.9, 19.9,
20.9, 22.0, 23.0, 24.0, 25.0, and 26 MeV,
there are available 
measurements separately for n$_0$ and n$_1$ by Poppe {\em et al.} [12].
Above 26 MeV, there are fewer experimental data at only a limited
number of incident energies and emission angles, and furthermore, only
data for the sum of n$_0$ and n$_1$ are available.  We used the
measurements by Batty {\em et al.} 
[22] at 30 and 50 MeV, by
Goulding {\em et al.} 
[23] at 119.8 MeV, and by Watson
{\em et al.} 
[24] at 134.2 MeV.  We fitted these data using
for the ratio $R = n_1 / n_0$ a fixed value of 0.35, as measured by
Poppe {\em et al.} 
[12] at $E_p = 25.9 MeV$, and smoothly 
interpolated/extrapolated the resulting Legendre coefficients 
 up to $E_p = 150$ MeV.  

As an example, Fig. 2 shows the sum of n$_0$ and
n$_1$ measured total production cross section 
(for references, see 
[5]) as a function of proton
kinetic energy, $E_p$, compared with our evaluation. One can see
good agreement in the whole energy range up to 150 MeV.

\begin{figure}[h!]
\vspace*{-0.5cm}
\centerline{
\hspace*{+0.5cm}
\psfig{figure=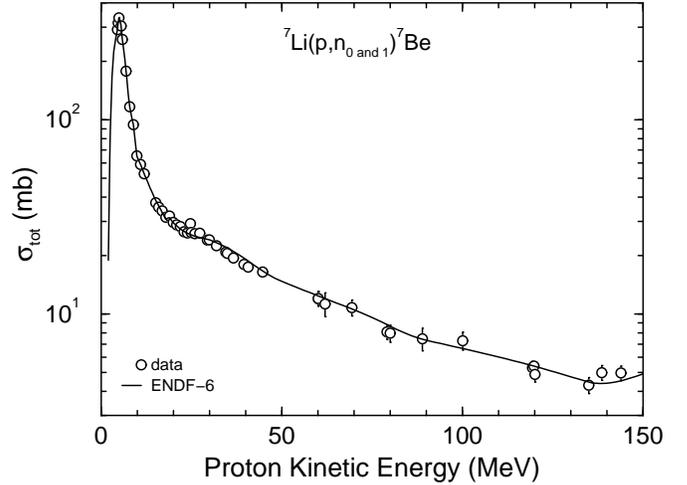,width=80mm,angle=-90}}

\caption
{
Comparison of the sum of
measured [12, 15, 16, 24, and 25]
laboratory angle integrated
cross sections for the 
$^7$Li(p,n$_0)^7$Be
and
$^7$Li(p,n$_1)^7$Be reactions (for details, see [5])
with our evaluation in the ENDF-6 format.  
}  
\end{figure}
	
\begin{center}
B. GNASH calculations
\end{center}

The latest version of the GNASH code has been described in Ref. [20],
and its latest application in nuclear reaction 
evaluation work has been described in Ref. [4].
For this reason, here we provide only an 
overview of the models used in the calculations,
concentrating on new features.

GNASH calculations of preequilibrium and Hauser-Feshbach decay require
input parameter information describing the optical model transmission
coefficients, nuclear level densities, gamma-ray strength functions, 
as well as nuclear level information for all the nuclei that can 
be populated in the reaction. For most of these quantities, we 
used default parameter information, as described in Ref. [4].
Below, we provide information on the optical model we used for 
$^7$Li.

Although the appropriateness of an optical potential for nucleon
scattering on a light nucleus such as lithium is questionable, we
have, for pragmatic reasons, made use of a relativistic potential for
$^6$Li published by Chiba {\it et al.} [26].
This potential was shown to provide a 
surprisingly good representation of available elastic scattering 
and total cross section data, for energies up to a few hundred 
MeV [26].
We converted this neutron potential 
for use on $^7$Li by making 
small isospin corrections, and produced a version for proton 
scattering by including Coulomb correction terms. The 
modified
potentials resulted in a calculated total neutron cross section that 
accounted for measured data fairly well. The calculated proton reaction 
cross section was also in good agreement with measurements (see 
Fig. 3).

\begin{figure}[h!]
\vspace*{-0.5cm}
\centerline{
\hspace*{+0.4cm}
\psfig{figure=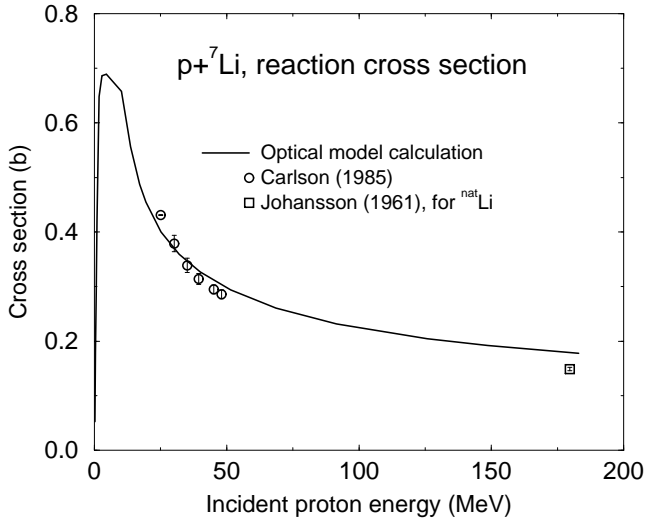,width=78mm,angle=-90}}

\caption
{Proton reaction cross section of $^7$Li. The solid curve shows
values calculated with the optical model by Chiba {\em et al.} [26]
modified here as described in Section III.B. Experimental data are
from Refs. [27, 28].
}  
\end{figure}

\vspace*{0.5cm}
Since the GNASH code includes n$_0$ and n$_1$ neutron emission contributions 
from compound nucleus and preequilibrium decay, these cross sections 
had to be ignored since, in our evaluation, they are based on 
measurements. Therefore we wrote a utility code that takes a 
GNASH output, modifies the  n$_0$ and n$_1$ neutron emission cross 
sections to zero, and in the evaluated file introduces the cross 
sections based on experiment as described in Sec.~III~A above.

Examples of the $^7$Li(p,xn) calculated zero-degree
double-differential cross sections of neutrons are shown in Fig.~4,
compared with measurements. Solid lines show the results from the
GNASH calculations; Dashed-lines show the n$_0$ and n$_1$
contributions (evaluated from the measurements). Comparisons are made
with $E_p = 15$, 20, and 30 MeV data of McNaughton et
al. [13],
$E_p = 40$ MeV data of Jungerman {\it et al.} [14],
and at $E_p = 55$, 90, and 140 MeV data of
Byrd and Sailor [29].
In these figures, in addition to the
highest-energy peak due to n$_0$ and n$_1$ neutrons, several
lower-energy peaks due to n$_2$ and n$_3$ neutrons, representing the
population of the 4.55- and 6.51-MeV states in $^{7}$Be are evident
(see also, e.g., Fig. 2 in Poppe {\it et al.} [12],
where even the
peaks by n$_4$ and n$_5$, representing the population of the 7.19- and
10.79-MeV states in $^7$Be are clearly observed). 

\begin{figure*}[h!]
\vspace*{-3.5cm}
\centerline{
\psfig{figure=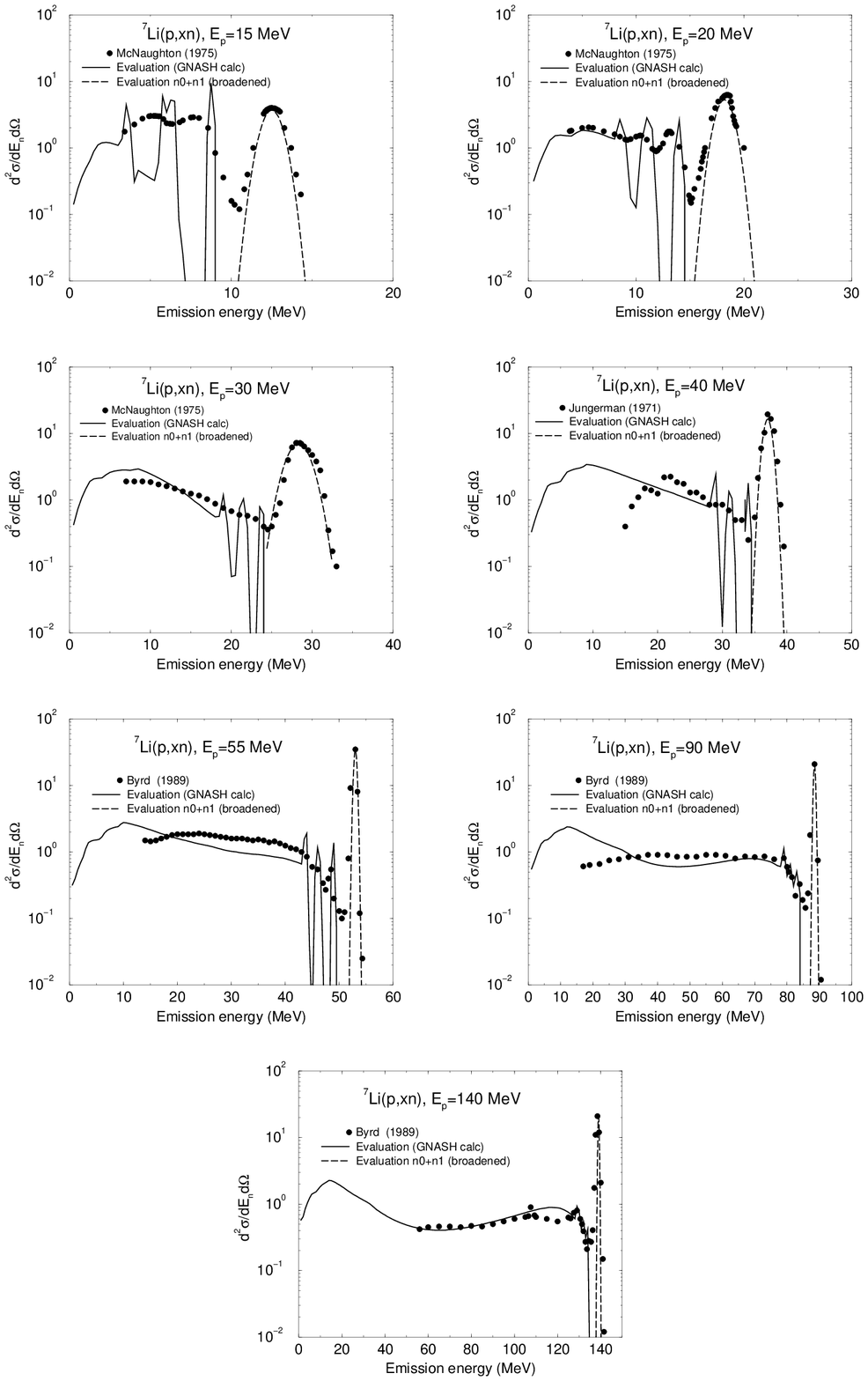,width=180mm,angle=-0}}

\vspace*{-1cm}
\caption
{
Zero degree neutron emission from the $^7$Li(p,xn) reaction.
Solid lines show the results from the GNASH calculations,
dashed lined show n$_0$ and n$_1$ contributions evaluated from
the measurements.
The GNASH-calculated spectra have not been broaded using the experimental
resolution.
Data are from Refs. [13, 14, 29].
}  
\end{figure*}

We emphasize that the GNASH results shown in Fig.~4 have not been 
broadened to account for the experimental detector resolution. This 
explains why peaks and dips are observed in the calculated energy-dependent 
spectra, but are not seen so prominantly in the experimental data. 
Overall, it is evident that the evaluation accurately represents 
the  n$_0$ and n$_1$ neutrons, as it should since it is based 
on the experimental data. The lower-energy neutrons from the 
GNASH calculation agree with the measurements less well, but 
it is hoped that the accuracy obtained is sufficient for 
LEDA design calculations.

\vspace{0.4cm}
\noindent IV. ENDF DATA
\vspace{0.1cm}

In this section we show some graphical representations of the
evaluated ENDF data. These figures help illustrate certain features of
the evaluation. The figures shown here are a small subset of figures
produced automatically by the NJOY code [30].
The full set of figures can
be viewed on the T-2 WWW site, by first going to
http://t2.lanl.gov/data/he.html, registering, and then clicking on
premade plots in Postscript format for protons       
(http://t2.lanl.gov/data/p-ps/).

Figure 5 shows the proton elastic scattering angular distribution for
incident energies up to 150 MeV. The increased forward-peaking 
with increasing incident energy is evident. Figure 6 shows the 
angular distributions, reconstructed from the ENDF Legendre coefficients,
for the n$_0$ and n$_1$ emitted neutrons.

Four separate figures are combined together in Figure 7. On the top left, 
the overall cross sections for the n$_0$, and n$_1$ reactions are shown, as 
is also the ``remaining'' cross section, designated as MT=5, that is 
used in the GNASH calculations. The sum of all these cross sections is
the reaction cross section as shown in Fig.~3.

In Fig.~7, top right, the inclusive production cross sections are shown 
for the light ejectiles. Note that, unlike for reactions on heavier 
nuclei, these do not keep increasing with incident energy up to 150 
MeV, but instead become approximately constant above 40 MeV. This
energy corresponds, approximately, to the total binding energy of $^7$Li, and
therefore above this energy it is not possible to obtain additional
particle production by increasing the incident energy.

In Fig.~7 lower-left, a perspective plot is shown of the angle-integrated 
neutron emission spectra obtained from GNASH, as a function of 
incident energy. The increasing role of preequilibrium high-energy 
neutron emission is seen as the incident energy increases. Finally, 
in Fig.~7 lower-right the total heating (MeV/collision) is shown as a function 
of incident energy. This includes energy transferred to all secondary 
charged-particles, except protons (since this is the projectile, and 
it is therefore assumed that a transport calculation will always be 
explicitly tracking the secondary protons).

\vspace{0.4cm}
\noindent V. SUMMARY AND FUTURE WORK
\vspace{0.1cm}

Our new evaluation for protons on $^7$Li is available in the ENDF-6 
format for use in radiation transport calculations. By using and testing  
these data in simulations of thick and thin lithium targets, we hope
to obtain feedback on possible improvements that are needed. 

An example of a recent test [3] of our library is shown in Fig. 8.
We have modeled the 43-MeV proton source from JAERI's TIARA AVE cyclotron
with MCNPX. Protons impinge on a 3.6-mm thick $^7$Li target. Resulting 
neutrons are constrained by an iron collimator 10.9 cm in diameter and 225
cm long. We have modeled this target, assuming a monoenergetic point 
proton source. The neutron flux is tallied on the surface exiting
the collimator. In the report with experimental results [31],
the neutron flux is normalized to unity in an energy band
between 36.3 and 45.5 MeV. We have normalized our results in the same
manner. Two MCNPX calculations were performed, one using tables by
our library and the other using the Bertini Intranuclear
Cascade Model (INC) in conjunction with a preequilibrium+evaporation
model.

Results for the experiment and the two calculations are shown in Fig. 8.
Overall, neither calculation is in completely satisfactory agreement
with the experiment. This is not surprising given 
we used a cascade, a preequilibrium, and an evaporation model
in the Bertini INC calculations
and the physics of the GNASH code
for such a light target.
However, we observe that the width of the neutron peak
more closely matches the experiment
when proton tables from our library are used.
This could be expected in advance, since we used for the peak
a parametrization of experimental data in our tables. 
The Bertini INC predicts a neutron peak that 
is lower in
energy and much broader than observed.

\begin{figure*}[h!]

\vspace*{-11cm}
\centerline{
\psfig{figure=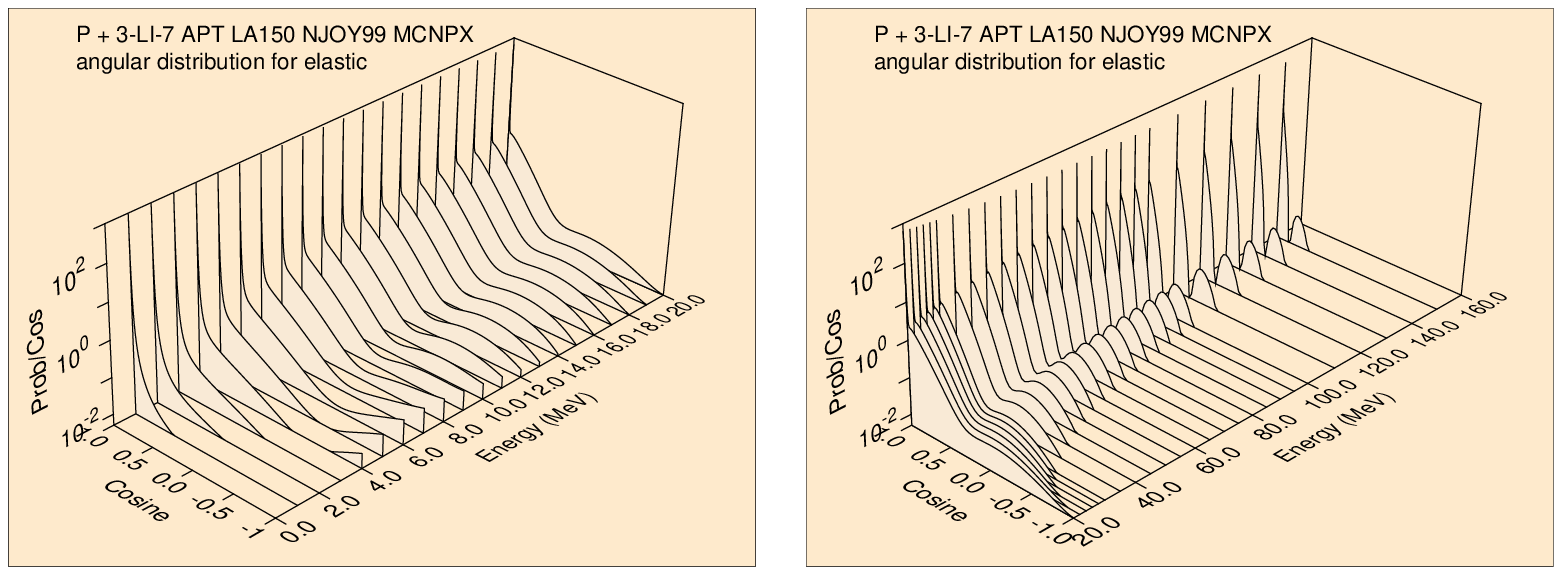,width=190mm,angle=-0}}

\vspace*{-8.5cm}
\caption
{
Proton elastic scattering angular distribution of $^7$Li at
different incident energies up to 150 MeV.
}  
\end{figure*}
									  
%\newpage

\begin{figure*}[h!]

\vspace*{-10cm}
\centerline{
\psfig{figure=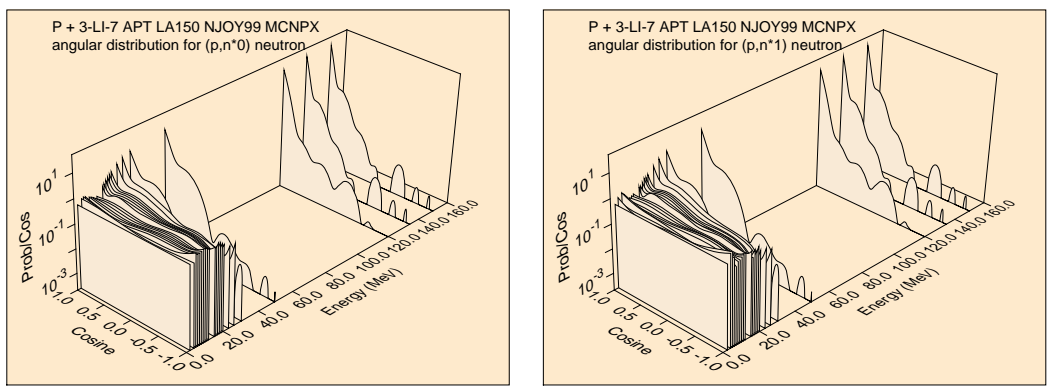,width=190mm,angle=-0}}

\vspace*{-8.5cm}
\caption
{
Angular distributions for n$_0$ and n$_1$ neutrons from p + $^7$Li
interactions at 
different incident energies up to 150 MeV.
}  
\end{figure*}

\begin{figure*}[h!]

\vspace*{-6cm}
\centerline{
\psfig{figure=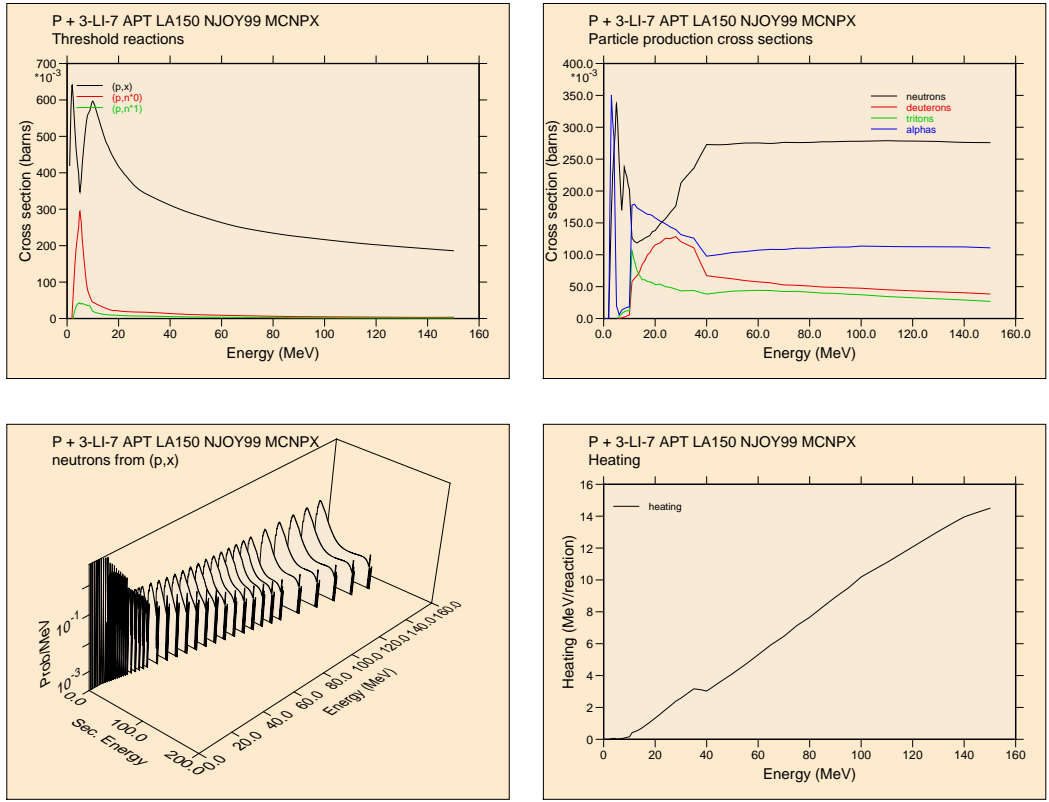,width=190mm,angle=-0}}

\vspace*{-5.5cm}
\caption
{
Different channel cross sections, neutron angle-integrated energy spectra,
and heating (MeV/collision) as functions of incident energy for
p + $^7$Li interactions, as indicated.
}  
\end{figure*}

\begin{figure*}[h!]

\vspace*{-2cm}
\centerline{
\hspace*{3mm}
\psfig{figure=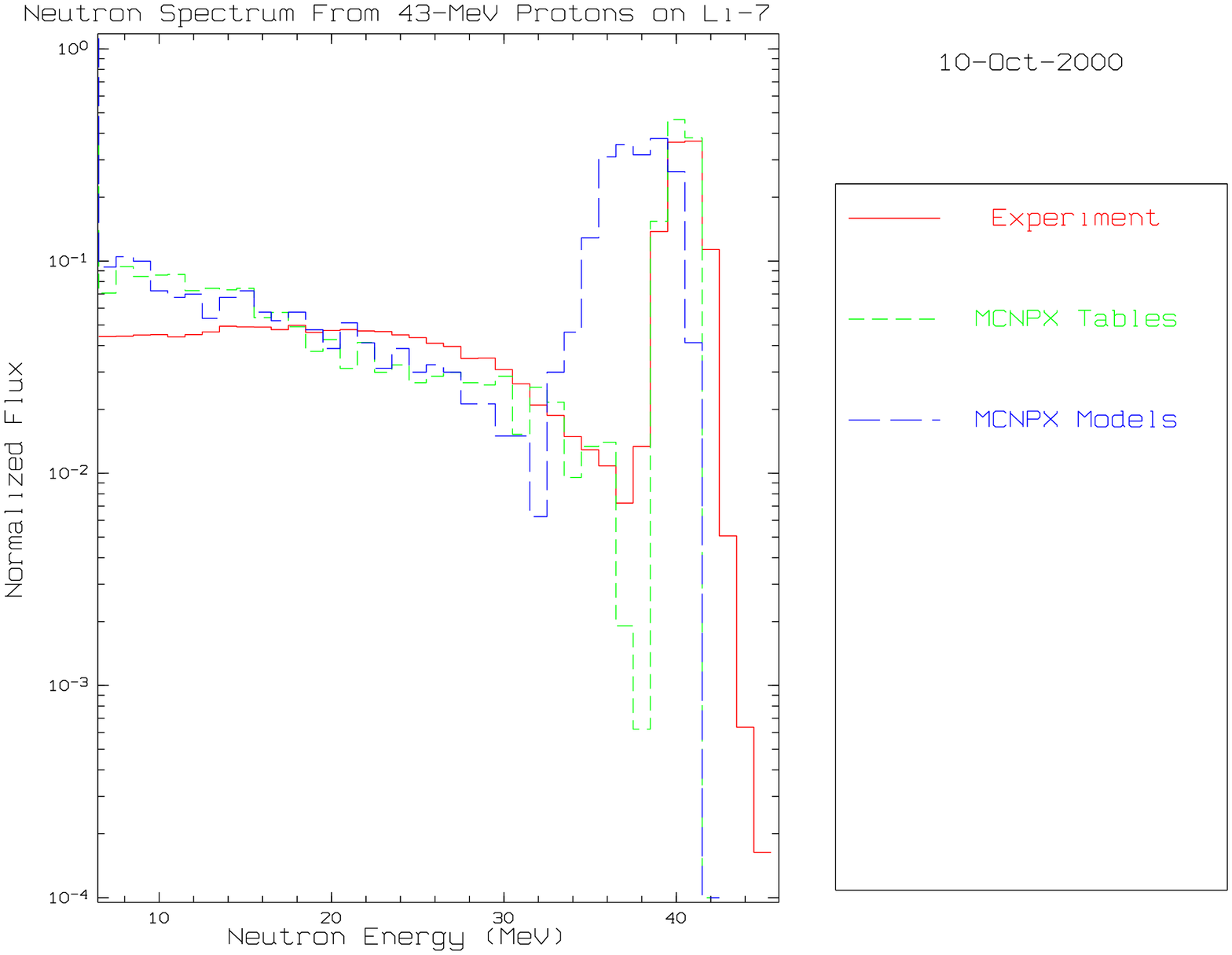,width=150mm,angle=-90}}

\vspace*{-1.0cm}
\caption
{
Neutron flux from a 3.6-mm thick $^7$Li target normalized to unity in the
energy band between 36.3 and 45.5 MeV calculated with MCNPX using
our data library (dashed histogram) and results by the Bertini INC
in conjunction with a preequilibrium+evaporation model 
(long-dashed histogram) compared with the experimental data from
[31] (solid histogram).
}  
\end{figure*}

For completeness, we list here a couple of areas where we know there
is room for improvement. The first point emphasizes
weaknesses in the ENDF evaluation that may be problematic for
simulations above 30 MeV for our case (see details in [5]).

Secondly, our analysis was not developed to include an accurate
treatment of gamma-ray production, but instead focussed on neutron
emission. Indeed, the nuclear level information in the GNASH
calculations describing the decay of excited states by gamma-ray emission 
was not carefully checked. Therefore, if gamma-ray emission becomes 
of concern in LEDA target design studies, an upgrade to this 
evaluation may be needed.

\newpage
\noindent ACKNOWLEDGMENTS
\vspace{0.1cm}

The authors are grateful to J.~L.~Ullmann and W.~B.~Wilson 
for useful discussions and information.

This study was supported by the U.~S.~Department of Energy.

\vspace{0.5cm}
\noindent REFERENCES
\vspace{0.1cm}
\begin{enumerate}

\vspace*{-0.2cm}
\item 
J. C. Browne, J. L. Anderson, M. W. Cappiello, G. P. Lawrence, and
P. W. Lisowski,
``Status of the Accelerator Production of Tritium (APT) Project,"
{\em Proc. APT Symp. The Savannah River Accelerator Project and 
Complementary Spallation Neutron Sources}, 
University of South Carolina, Columbia, USA, May 14-15, 1996, p. 14,
F. T. Avignone and T. A. Gabriel, Eds.,
World Scientific, Singapore, (1998).

\vspace*{-0.2cm}
\item 
{\em ED\&D Monthly Report, October, 1999},
LA-UR 99-6201, Los Alamos National Laboratory (1999). 

\vspace*{-0.2cm}
\item 
H. G. Hughes, M. B. Chadwick, R. K. Corzine, H. W. Egdorf,
F. X. Gallmeier, R. C. Little, R. E. MacFarlane,
S. G. Mashnik, E. J. Pitcher, R. E. Prael, A. J. Sierk,
E. C. Snow, L. S. Waters, M. C. White, and P. G. Young,
``Status of the MCNPX Transport Code,"
Los Lalmos National Report LA-UR-004942,
Proc. Int. Conf. on Advanced Monte Carlo for Radiation Physics,
Particle Transport Simulation and Application (MC 2000),
23-26 October, 2000, Lisbon, Portugal.

\vspace*{-0.2cm}
\item 
M. B. Chadwick, P. G. Young, S. Chiba, S. Frankle, G. M. Hale,
H. G. Hughes, A. J. Koning, R. C. Little, R. E. MacFarlane, 
R. E. Prael, and L. S. Waters,
``Cross Section Evaluationsto 150 MeV for Accelerator-Driven System 
and Implementation in MCNPX,"
{\it Nucl. Sci. Enf.}, {\bf 131}, 293 (1999).

\vspace*{-0.2cm}
\item 
S. G. Mashnik, M. B. Chadwick, P. G. Young, R. E. MacFarlane, 
and L. S. Waters,
``$^7$Li(p,n) Nuclear Data Library for Incident Proton Energies to 150 MeV,"
Los Alamos National Laboratory Report LA-UR-00-1067, Los Alamos (2000).

\vspace*{-0.2cm}
\item 
M. Drosg, 
``Sources of Variable Energy Monoenergetic Neutrons for
Fusion-Related Applications," 
{\em Nucl.~Sci.~Eng.},~{\bf 106}, 279 (1990).

\vspace*{-0.2cm}
\item 
J.~H.~Gibbons and H.~W. Newson, 
``The $^7$Li(p,n)$^7$Be Reaction," 
{\em Fast Neutron Physics}, Vol. 1, 
J.~B.~Marion and J.~L.~Fowler, Eds., Interscience, New York (1960).

\vspace*{-0.2cm}
\item 
J.~H.~Langsdorf,~Jr., J.~E.~Monahan, and W.~A.~Reardon, 
``A Tabulation of Neutron Energies from Monoenergetic Protons on Lithium," 
ANL-5219, Argonne National Laboratory (Jan. 1954). 

\vspace*{-0.2cm}
\item 
J.~W.~Meadows and D.~L.~Smith, 
``Neutron Source Investigations in Support of 
the Cross Section Program at the Argonne Fast-Neutron Generator,"
ANL/NDM-53, Argonne National Laboratory (May 1980). 

\vspace*{-0.2cm}
\item 
J.~W.~Meadows and D.~L.~Smith, 
``Neutrons from Proton Bombardment of Natural Lithium," 
ANL-7938, Argonne National Laboratory (June 1972). 

\vspace*{-0.2cm}
\item 
J.~D.~Anderson, C.~Wong, and V.~A.~Madsen, 
``Charge Exchange Part of the Effective Two-Body Interaction," 
{\em Phys. Rev. Lett.}, {\bf 24}, 1074 (1970).

\vspace*{-0.2cm}
\item 
C.~H. Poppe, J.~D. Anderson, J.~C. Davis, S.~M. Grimes, and C. Wong,
``Cross Sections for the $^7$Li(p,n)$^7$Be Reaction Between 4.2 and 26 MeV,"
{\em Phys. Rev. C}, {\bf 14}, 438 (1976).

\vspace*{-0.2cm}
\item 
M.~W. McNaughton, N.~S.~P. King, F.~P. Brady, J.~L. Romero, and 
T.~S. Subramanian,
``Measurements of $^7$Li(p,n) and $^9$Be(p,n) Cross Sections at 15, 20, 
and 30 MeV," 
{\em Nucl. Instr. Meth.}, {\bf 130}, 555 (1975).

\vspace*{-0.2cm}
\item 
J.~A. Jungerman, F.~P. Brady, W.~J. Knox, T. Montgomery, M. R. McGie,
J.~L. Romero, and Y. Ishizaki, 
``Production of Medium-Energy Neutrons from Proton Bombardment of Light 
Elements," 
{\em Nucl. Instr. Meth.}, {\bf 94}, 421 (1971).

\vspace*{-0.2cm}
\item 
S.~D. Schery, L.~E. Young, R.~R. Doering, S.~M. Austin, and R.~K. Bhowmik,
``Activation and Angular Distribution Measurements of 
$^7$Li(p,n)$^7$Be ($0.0+0.429$ MeV) for $E_p = 25-45$ MeV: A Technique for 
Absolute Neutron Yield Determination," 
{\em Nucl. Instr. Meth.}, {\bf 147}, 399 (1977).

\vspace*{-0.2cm}
\item 
T.~E. Ward, C.~C. Foster, G.~E. Walker, J. Rapaport, and C.~A. Goulding,
``$1/E$ Dependence of the $^7$Li(p,n)$^7$Be (g.s.$ + 0.43$ MeV) Total Reaction 
Cross Section," 
{\em Phys. Rev. C}, {\bf 25}, 762 (1982).

\vspace*{-0.2cm}
\item 
T.~N. Taddeucci, W. P. Alford, M. Barlett, R. C. Byrd, T. A. Carey,
D. E. Ciskowski, C. C. Foster, C. Gaarde, C. D. Goodman, C. A. Goulding,
J. B. McClelland, D. Prout, J. Rapaport, L. J. Rybarcyk, W. C. Sailor,
E. Sugaebaker, and C. A. Whitten, Jr.,
``Zero-Degree Cross Sections for the $^7$Li(p,n)$^7$Be
(g.s.$ + 0.43$ -- MeV) Reaction in the Energy Range 80-795 MeV," 
{\em Phys. Rev. C}, {\bf41}, 2548 (1990).

\vspace*{-0.2cm}
\item 
M. Baba, Y. Nauchi, T. Iwasaki, T. Kiyosumi, M. Yohioka, S. Matsuyama,
N. Hirakawa, T. Nakamura, Su. Tanaka, S. Meigo, H. Nakashima,
Sh. Tanaka, N. Nakao, 
``Characterization of 40-90 MeV $^7$Li(p,n) Neutron Source at TIARA 
Using a Proton Recoil Telescope and a TOF Method," 
{\em Nucl. Instr. Meth. A}, {\bf 428}, 454 (1999).

\vspace*{-0.2cm}
\item 
V. McLane,
``ENDF-102 Data Formats and Procedures for the Evaluated Nuclear Data 
File ENDF-6,"
BNL-NCS-44945, Rev. 2/97, Brookhaven National Laboratory,
National Nuclear Data Center (1997).

\vspace*{-0.2cm}
\item 
P. G. Young, E. D. Arthur, and M. B. Chadwick,
``Comprehensive Nuclear Model Calculations: Theory and Use
of the GNASH Code,"
{\em Proc. IAEA Workshop on Nuclear Reaction Data and Nuclear Reactors
--- Physics, Design, and Safety}, Trieste, Italy, April 15 --
May 17, 1996, p. 227, A. Gandini and G. Reffo, Eds., 
World Scientific Publishing, Ltd., Singapore (1998).

\vspace*{-0.2cm}
\item 
P. G. Young, E. D. Arthur, and M. B. Chadwick,
``Comprehensive Nuclear Model Calculations: Introduction to the Theory and
Use of the GNASH Code,"
LA-12343-MS, Los Alamos National Laboratory (1992).

\vspace*{-0.2cm}
\item 
C. J. Batty, B. E. Bonner, A. I. Kilvington, C. Tschal\"{a}r, and
L. E. Williams,
``Intermediate Energy Neutron Sources,"
{\em Nucl. Instrum. Methods}, {\bf 68}, 273 (1969).

\vspace*{-0.2cm}
\item 
C. A. Goulding, M. B. Greenfield, C. C. Foster, T. E. Ward, J. Rapaport,
D. E. Bainum, and C. D. Goodman, 
``Comparison of the $^{12}$C(p,n)$^{12}$N
and $^{12}$C(p,p$'$) Reactions at $E_p = 62$ and 120 MeV,"
{\em Nucl. Phys. A}, {\bf 331} 29 (1979). 

\vspace*{-0.2cm}
\item 
J. W. Watson, R. Pourang, R. Abegg, W. P. Alford, A. Celler, S. El-Kateb,
D. Frekers, O. H\"{a}usser, R. Helmer, R. Henderson, K. Hicks, 
K. P. Jackson R. G. Jeppesen, C. A. Miller, M. Vetterli, S. Yen,
and C. D. Zafiratos, 
``$^7$Li(p,n)$^7$Be and $^{12}$C(p,n)$^{12}$N
Reactions at 200, 300, and 400 MeV," 
{\em Phys. Rev. C}, {\bf 40}, 22 (1989).

\vspace*{-0.2cm}
\item 
J. W. Watson, B. D. Anderson, A. R. Baldwin, C. Lebo, B. Flanders,
W. Pairsuwan, R. Madey, and C. C. Foster, 
``A Comparison of Methods for Determining Neutron Detector Efficiencies 
at Medium Energies,"
{\em Nucl. Instrum. Methods}, {\bf 215}, 413 (1983).

\vspace*{-0.2cm}
\item 
S. Chiba, K. Togasaki, M. Ibaraki, M. Baba, S. Matsuyama, N. Hirakawa, 
K. Shibata, O. Iwamoto, A. J. Koning, G. M. Hale, and M. B. Chadwick, 
``Measurements and Theoretical Analysis of Neutron Elastic Scattering and
Inelastic Reactions Leading to a Three-Body Final State for $^6$Li at
10 to 20 MeV,"
{\em Phys. Rev. C}, {\bf 58}, 2205 (1998).

\vspace*{-0.2cm}
\item 
R. F. Carlson, A. J. Cox, T. N. Nasr, M. S. De Jong, D. L. Ginther,
D. K. Hasell, A. M. Sourkes, W. T. H. Van Oers, and D. J. Margaziotis,
``Measurements of Proton Total Reaction Cross Sections for  $^6$Li,
$^7$Li, $^{14}$N, $^{20}$Ne, and $^{40}$Ar Between 23 and 49 MeV,"
{\em Nucl. Phys. A}, {\bf 445}, 57 (1985).

\vspace*{-0.2cm}
\item 
A. Johansson, U. Svanberg, and O. Sundberg,
``Total Nuclear Reaction Cross Sections for 180 MeV Protons,"
{\em Arkin f\"ur Fisik}, {\bf 19}, 527 (1961).

\vspace*{-0.2cm}
\item 
R. C. Byrd and W. C. Sailor, 
``Neutron Detection Efficiency for NE213 and 
BC501 Scintilators at Energies Between 25 and 200 MeV," 
{\em  Nucl. Instrum. Methods A}, {\bf 274}, 494 (1989).

\vspace*{-0.2cm}
\item 
R. E. MacFarlane,
``Recent Progress on NJOY,"
LA-UR-96-4688, Los Alamos National Laboratory (1996); 
the last version of the code, NJOY99, is described on the Web
page at hppt://t2.lanl.gov/codes/codes.html.

\vspace*{-0.2cm}
\item 
N. Nakao, H. Nakashima, T. Nakamura, S. Tanaka, S. Tanaka, K. Shin, M. Baba,
Y. Sakamoto, and Y. Nakane,
``Transmutation Through Shields of Quasi-Monoenergetic
Neutrons Generated by 43-MeV and 68-MeV Protons -- I: Concrete
Shielding Experiment and Calculation for Practical Application,"
{\em  Nucl. Sci. Eng.}, {\bf 124}, 228 (1996).

\end{enumerate}

\end{document}